\documentclass[article,12pt]{JHEP3}
\usepackage{graphicx}
\usepackage{epsf,colordvi,shadow,pifont}
\usepackage{amsmath,epsfig}
\usepackage{amssymb,amsfonts}
\usepackage{latexsym}
\usepackage{slashed}
\usepackage{epsfig}
\usepackage{showkeys}
\usepackage[vcentermath,enableskew]{youngtab}
\def\hri#1#2{\href{http://arxiv.org/abs/#1}{[ArXiv:#1]#2}}
\def\hre#1#2{\href{http://arxiv.org/abs/#1/#2}{[ArXiv:#1/#2]}}
\def\hspi#1#2{\href{http://www.slac.stanford.edu/spires/find/hep/www?irn=#1}{#2}}
\relax

\renewcommand{\theequation}{\arabic{section}.\arabic{equation}}
\newcommand{\bb}[1]{\bar{\textbf{#1}}}
\def\be{\begin{equation}}
\def\ee{\end{equation}}

\newcommand{\la}{\lambda}

\newcommand{\bear}{\begin{eqnarray}}
\newcommand{\bea}{\begin{eqnarray}}
\newcommand{\eear}{\end{eqnarray}}
\newcommand{\eea}{\end{eqnarray}}
\def\hre#1#2{\href{http://arxiv.org/abs/#1/#2}{[ArXiv:#1/#2]}}
\def\hspi#1#2{\href{http://www.slac.stanford.edu/spires/find/hep/www?irn=#1}{#2}}
\newbox\pippobox

\renewcommand{\b}[1]{\textbf{#1}}

\def\II{\relax{\rm I\kern-.18em I}}

\def\e{\epsilon}
\def\m{\mu}
\def\n{\nu}
\def\r{\rho}
\def\s{\sigma}
\def\pa{\partial}

\def\sp{\;\;\;,\;\;\;}

\def\a{\alpha}
\def\b{\beta}

\def\tr{\ensuremath{\mathrm{Tr}}}
\def\gn{{\rm G_N}}

\def\g{\gamma}
\def\d{\delta}

\title{Gravity and axions from a random UV QFT.}

\author{\large Elias Kiritsis\\
 ~\\
 \href{http://hep.physics.uoc.gr/}
 {Crete Center for Theoretical Physics}, Department of Physics, University of Crete
 71003 Heraklion, Greece\\
  \centerline{and}\\
\href{http://www.apc.univ-paris7.fr}
{APC, Universit\'e Paris 7}, CNRS/IN2P3, CEA/IRFU, Obs. de Paris, Sorbonne Paris Cit\'e, B\^atiment Condorcet, F-75205, Paris Cedex 13, France (UMR du CNRS 7164).\\
 \centerline{and}\\
\href{http://wwwth.cern.ch/}{Theory Group, Physics Department, CERN}, CH-1211, Geneva 23, Switzerland }

\preprint{CCTP-2014-14\\CERN-PH-TH/2014-153}

\abstract{It is postulated that the UV QFT is enormous and random. The coupling of the Standard Model to such QFT is analyzed. It is argued that massless 4d gravity and axions are general avatars of the postulate. The equivalence principle emerges naturally as well as a concrete set of sources for its breaking. The axion scale is related to the 4d Planck scale as $f=M_P/N$, where $N$ is the ``number of colors" of the (almost) hidden UV CFT.    }


\begin{document}

\section{Introduction}
\setcounter{equation}{0}

Gravity is the first interaction to be successfully described in modern science. It remains todate the least understood.

There are two main reasons that  set gravity apart from the other known interactions. The first is the fact that the  classical theory of general relativity is at best an effective field theory as a quantum field theory (QFT), giving no useful clues on its UV completion.
The second is that there is a major clash between our understanding of QFT and gravity summarized in the cosmological constant
problem. This was made sharper by the recent observation of the ``dark energy" filling the universe.

Along with the problems came new ideas on the structure and UV completion of gravity. String theories provide a perturbative
quantum completion of the gravity theory, introducing a new intermediate scale, the string scale. It is a well controlled theory once energy transfers  remain  much smaller than  the Planck scale.
Another comes from large-N holographic ideas that relate  adjoint large-N theories to string theories (containing gravity).

Perturbative string theory cannot describe high-energy gravitational physics once energy transfers are of the order of the Planck scale, as perturbation theory breaks down.
It would seem that non-perturbative dualities might give a way
out, since they provide information about strong coupling physics.
Indeed non-perturbative dualities relate theories with different
(dimensionless) couplings and string scales.
This is however not the case for the gravitational interaction, since any
non-perturbative duality we know, leaves the Planck scale fixed,
and thus cannot address questions on physics at or beyond the
Planck scale.
One is therefore drawn to find other ideas that will provide clues on the nature of the gravitational physics at short distances.

It was suggested in section 7.8 of \cite{rev} and elaborated further in \cite{talk} that all interactions in
 nature are described by four-dimensional quantum field  theories.
In particular observable gravity is the avatar of a ("hidden"
 in the IR)  large-N gauge theory,
that couples to SM matter via bifundamental fields. Moreover,
a geometrization of this picture suggests the existence of a
(Copernican) chaotic brane universe. In such a universe the SM is represented by an appropriate  local stack of
 branes embedded in a higher dimensional universe filled with
 other branes. The relative  motion of branes is responsible
to a large degree for the ``local cosmology" observed in our universe, \cite{mirage}.
In a sense, this postulate includes a new level of structure in the cosmos, as well as ``extra dimensions". Such dimensions beyond the four observed ones  are emergent in this picture.

In this postulate, gravitons are composite at high energies and their constituents are ``large-N gluons". Today such a postulate raises few eyebrows due to the Maldacena conjecture \cite{mald1,mald2} and subsequent works in the last two decades.

Since the early work of 't Hooft \cite{hoof} it was understood that
the low-energy limit of large N-gauge theories is described by some string theory.
The gauge theory versus string theory/gravity
correspondence is a more precise indication that gravity can be realized as an
effective theory of a four-dimensional gauge theory. The inverse
is also true: fundamental string theory in some backgrounds
describes the physics of theories that at low energy are standard
gauge theories. Although the bulk-boundary duality is a concept
transcending that of four-dimensional gauge theories, it is most
powerful in the four-dimensional cases.

The lesson of the AdS/CFT correspondence is that any
gauge theory has a dual gravity/string theory.
The idea of 't Hooft that gravity must be holographic \cite{holog1}
indicates that a gravity theory must have a dual gauge theory
description.\footnote{We are using the term gauge-theory in a loose sense to include adjoint and bifundamental QFTs. There is a distinction between ``color" that stands for a gauge degree of freedom, and ``flavor" that stands for ungauged degrees of freedom.}

 A standard (confining) gauge theory realization of four-dimensional  gravity generically
predicts massive composite gravitons. The graviton is the spin-two glueball generated out of the vacuum by the stress-tensor of the theory. Confinement typically comes together with a mass gap.
A graviton mass is severely constrained by observations. Its presence may have two potential advantages.
It will generate an intrinsic  cosmological constant that may be of the
order of magnitude observed today, if the graviton mass modifies
gravity at or beyond the horizon today as described in section 7.7 of \cite{rev}. Also, the fact that the
graviton is a bound state, provides  a mechanism
for suppressing the observable cosmological constant.
In particular, the graviton {\sl does not} directly couple
to the standard ``vacuum energy" of the SM
fields.
Such ideas face non-trivial challenges, \cite{pol},  and it is clear if they are the final word.

The approach advocated in \cite{rev} has some similarities but also important differences with ideas in
\cite{sundrum1,sundrum2} and \cite{zee}.
The interplay between gauge theory and gravity was suggested by black hole thermodynamics considerations and gave a new view on the thermodynamics of gravity. This view is partly captured by Jacobson's observation that  GR dynamics is many cases is a consequence of the first law of thermodynamics, \cite{jacob},  Verlinde's recent reformulation of gravity as an entropic force, \cite{erik} and connections to entanglement entropy, \cite{myers}.

In this paper we will review the main elements of the proposal in \cite{rev,talk} and we will provide more details and some further thoughts that emanate from a Copernican version of the SM physics.

\section{The UV landscape}

The basic postulates that we will explore here are:

\begin{enumerate}

\item The theory of the world is a UV-complete 4d QFT.

\item Its size is enormous and its structure close to random.

\item The SM is a small sector of this QFT.

\end{enumerate}

The motivations for such postulates are varied and rely on certain principles

\begin{itemize}

\item  The Aristarchus-Copernicus principle: we are probably  not at the center of the ``world".

\item H. Nielsen's idea from the '80s that the UV QFT is random, \cite{hbn}.

\item The realization in the past two decades that the concept of a hidden sector in QFT and string theory is not only a possibility but a necessity.

\item The gauge/gravity correspondence that provided a fresh look both at gauge theories and the gravitational/string forces.

\item  The realization that string theory, if it is a unique theory has most probably an  enormous landscape of ``vacua" with many different properties, \cite{land}.

\end{itemize}

 One main motivation to entertain these postulates is in order to generate gravity as an emergent force and make the gravitons low-energy excitations that are composite at high energy.
Another is that the general structure of QFT as we understand it today makes natural and generic the effect that low energy QFTs can be hidden from others by the high mass of ``messenger" degrees of freedom or the weakness of associated interactions. This is a state of affairs that is ubiquitous in theories with a landscape of vacua like string theory.

The main postulate is therefore that there is an enormous number of fields in the four-dimensional QFTs at high energy. Moreover the structure of the UV theory and its interactions is ``generic" with some distribution of its properties like size of gauge groups, associated couplings etc.

An important ingredient in this is that this theory corresponds to a UV-complete,  QFT, namely all couplings in the extreme UV are either asymptotically free or conformal.
This requirement of UV-completeness poses a serious set of constraints on the UV theory.

A first set of important factors are the gauge group factors. The generic UV gauge group is a generic product of simple factors. To simplify the discussion we will take them to be $SU(N_i)$ factors. The ranks $N_i$ may range from a few to astronomical numbers of the type $10^{30}$ or more.
 As we will see further the dominant gauge groups that are visible at low energies are associated with continuum QFTs: they are conformal.

Therefore in the generic theory at weak coupling the spectra can be classified in terms of vectors, scalars and spin-${1\over 2}$ fermions.
A priori, apart from the vectors that transform in adjoints of gauge groups the other fields can transform in arbitrary representations. However, the Casimirs of generic representations increase typically much faster that that of the adjoint\footnote{This is obvious from the one-loop $\beta$-function in gauge theories and from the table of casimirs that can be found in the appendix of \cite{4dstring}.}.
Therefore, in order for the $\beta$ functions to be non-positive,  for gauge groups with large enough $N_i$, the only representations that will be allowed
are of the fundamental or bifundamental type (adjoint, $\Yboxdim8pt\yng(1,1)$ and $\Yboxdim8pt\yng(2)$).

Beyond weak coupling, similar statements are valid. Although one cannot speak directly about color representations, the total number of degrees of freedom must at most increase as ${\cal O}(N^2)$. This can be shown by bounding the degrees of freedom by those of AF theories which abide by the perturbative argument (at high energies) and therefore have at most  ${\cal O}(N^2)$ degrees of freedom.

 The rest would follow if any strongly-coupled 4d CFT can be thought of as the IR fixed point of a flow that originates in an AF UV Theory.
This however it is not known if it is true in 4d. It has been conjectured
 however by M. Douglas for 2d CFTs.
We know however that this is not any more true in six dimensions where CFTs can have an ${\cal O}(N^3)$ degrees of freedom and free CFTs have no relevant operators.

\subsection{Coupling different gauge group factors: the messenger sectors}

There are a priori two ways of coupling together different gauge group factors.

\begin{itemize}

\item  By messenger fields. They link two different gauge groups if they transform under both of them. They will be particularly important in communicating interactions between groups of large rank and  groups with finite rank like the SM.

 Such fields must be generically transforming in the $\Yboxdim8pt\yng(1)$ of SU(N) with $N\gg 1$ without further restrictions on the representation of the ``small group". We will call such fields the {\it fundamental messengers} from now on. Adjoints and  $\Yboxdim8pt\yng(1,1)$ and $\Yboxdim8pt\yng(2)$ are also in principle allowed by the $\beta$-function bounds, but they can only couple to very few small groups, and provided their rank is very small.
  We will call these fields the {\it exceptional messengers}.

\item  Multiple trace interactions.  Indeed different gauge groups can be coupled via double-trace (or higher-trace) deformations that are marginal or marginally relevant. A detailed analysis of such couplings in a large-N context was performed in \cite{n1,n2} along with an analysis of new fixed points coupling distinct CFTs.

    Such couplings can also be generated by the limit of messenger fields pushed to infinite mass. A comparison with the two pictures was presented in the concluding section of \cite{n1}. However, standard marginal couplings cannot be generated as they will violate the decoupling theorem.

\end{itemize}

An important issue concerns masses of fields that couple different simple gauge groups.For scalars in the generic case there are no natural finite masses except if they are protected by a symmetry (like supersymmetry) or are accidentally fine-tuned.
For fermions, masses can be anything provided they do not come from symmetry breaking associated with scalars (in which case the vevs will have a typical hierarchy problem). Similar remarks apply also to vectors. The natural case is that masses are due to vevs of scalars with flat potentials in CFTs that are not affected by the hierarchy problem.

Of interest in the sequel will be the theories that are coupled with messengers to the Standard model. Our main concern will be which theories  are such so that their effects on the SM are visible at arbitrarily low energies.

We will entertain first all the qualitative possibilities and then decide which is dominant.

We have the following choices for the hidden gauge group: Weak coupling versus strong coupling, large rank versus small rank, and light messengers versus heavy messengers. The notion of light or heavy is always decided by comparison to SM mass scales.
On top of these possibilities there is a further one, the presence or absence of a mass gap in the hidden theory, which is the choice between conformal invariance and asymptotic freedom.

Also we must separate two classes of bound states that will be eventually be coupled to the SM because of the existence of messengers.

\begin{itemize}

\item ``Hidden bound-states": they are the bound states of the hidden gauge theory, and they involve the standard glueballs, with among the universal ones, the $0^{++}$ (corresponding to $Tr[F^2]$), the $2^{++}$ (corresponding to stress tensor), and the $0^{-+}$ corresponding to the instanton density. They are not charged under the SM gauge group, and their couplings to SM particles will be induced by the messenger interactions.

\item ``Messenger bound-states": These involve meson-like bound-states of the messenger fundamentals, and could also include more operators of the hidden theory. They are non-trivially charged under the SM gauge group and a subset is  expected to be in one-to-one correspondence with the SM particles.

    \end{itemize}

It can be directly argued that if the hidden theory has a serious mass gap (that suggests also strong coupling in the IR) the hidden   bound states
with be heavy and therefore they will not be observable in the far IR.
If the messenger masses are heavier that the characteristic scale of the hidden theory,  then
the messenger bound-states (or ``messenger mesons") will also be heavy and therefore invisible in the IR.

If on the other hand the ``hidden" theory  is AF and the messengers are light then the only light degrees of freedom that could couple to the SM in the IR are the ``hidden pions".\footnote{A counterexample to this is discussed in appendix B of \cite{4dstring}, which however is not relevant in the present discussion. The issue there is an emergent IR strong coupling scale that can be exponentially smaller than the messenger masses. }
 This is indeed what is expected to happen in the technicolor paradigm, and this is an allowed possibility that could provide among other things electroweak breaking in the SM.
 It is important to stress that in this case, the relatively light composite degrees of freedom are ``charged" under the SM gauge group.

In the opposite case the hidden theory can be (approximately) conformal at low energy. In this case the presence of (very) light messengers is defacto forbidden by the definition of the SM model as the full set of low-lying fields with non-trivial quantum numbers under the SM gauge group. Therefore the messengers must have a large mass.

 Before moving further we summarize our conclusions. A hidden QFT coupled with messengers to the standard model will be visible in the IR if (a) it is AF and the messengers light (in that case it is in the technicolor class) (b) It is (nearly) conformal and the messengers are heavy.

 The only two other options to decode the nature of the hidden CFT  is whether it is weakly coupled or strongly coupled and whether it has a large or small rank.

 \subsection{The dominant 't Hooft coupling}

 We would like to argue first that the hidden CFT that is dominantly coupled to the SM at the far IR is a strongly coupled CFT. The main reason is stability. We will see this issue coming up in several places in this work.

 A stable theory must be free of relevant or marginally-relevant operators. Such operators if present destabilize the CFT. For example in the presence of messenger fields, such operators will be generated indirectly by  SM and messenger interactions and will create mass gaps in the IR screening in this way such hidden theories from the IR of the SM
Therefore, the successful theory must be free of such operators.
Weakly coupled (approximate) CFTs always have such relevant operators, namely masses for bosons and fermions. Chiral gauge theories evade this argument but they do not seem to exist at large rank and be conformal, as we will argue in the next subsection.

Supersymmetry, helps with stability but not all the way. The most supersymmetric of 4d CFTs, N=4 sYM is known to have several BPS protected relevant operators that can give masses to bosons and fermions.

Therefore the only chance to find stable CFTs is at strong coupling. In two dimensions it is  known that there are  CFTs without relevant operators, \cite{schellekens}. One can also find CFTs that lack also marginal operators.  It is  not known however  whether such CFTs exist in four dimensions, but this may be due to our limited list of four-dimensional CFTs.

\subsection{The dominant rank}

We would now like to argue that out of all strongly coupled CFTs coupled via messengers to the SM  the one that has the best chance of being scale invariant is the one that has the maximal rank. The idea is that one needs a large rank in order for the scale invariance not to be seriously perturbed by messengers.

A strongly coupled CFT with finite rank  is importantly and seriously perturbed when we add a set of messager fields that couple it to another finite-rank theory. The reason is that $\beta$ functions are not suppressed, and even though the spectrum of messengers is carefully chosen, (for example 6 scalars and four fermions per messenger vector) higher order terms in $\beta$ functions do not typically cancel, and the conformal symmetry, that is important for IR survival  is compromised.

A related issue is that vectors on one hand are important in balancing the $\beta$-function contribution of messengers, but in order for them to exist as massive particles they must be protected by a gauge symmetry. This implies that they are part of a spontaneously broken gauge group. As they are charged both under the hidden and the SM gauge group, then it natural to expect that if there are messenger vectors, there should be a phase where parts of the SM and the hidden gauge group are unified. In a sense this is an interesting variant of Grand unification, the main difference being that this is unification with gravity. We will call this phenomenon ``gravitational unification".

Fundamentals can drive AF theories to IR fixed points, a fact that could be acceptable for our purposes. This is the essence of the Banks-Zaks (BZ) argument, \cite{bz}.

We do know at least in supersymmetric cases that the conformal window extends well below the BZ region. In all known cases, one needs a number of fundamentals that is of the order of the number of colors.
If the characteristic scale of the hidden AF theory is $\Lambda_{\rm hidden}$ then the masses of the fundamentals $m_f$ must be much lighter: $m_f\ll \Lambda_{\rm hidden}$ in order for the fixed point to be approached.
This is analyzed in detail in appendix B of \cite{4dstring}, where it is shown that in the generic case of the conformal window, for light messenger masses $m_f$, the IR YM scale $\Lambda_{IR}$ generated is of the same order as the fundamental masses $m_f$. Therefore,   all bound-states, glueballs and ``pions" have masses of the same order as the messenger scale. For these to become massless the messengers must be almost massless, and we end up  in a previous case that we excluded.

An exception to this is if the number of colors and messengers are large and tuned so that we are in the BZ region, \cite{4dstring}. In this case, $\Lambda_{IR}\ll m$ exponentially in the BZ tuning. The neutral bound-states (glueballs) are much lighter than the mesons, who are charged under the gauge group.

 The large number of  messengers in this last context can be distributed over various spectators groups, one of which is the SM. If their masses are roughly of the same order of magnitude, and well below the BZ scale, then the picture above remains intact. The SM is coupled to the universal sector with exponentially small masses, while the mesons, charged under the SM are few with masses of the order of the messenger scale. In that case the rank of the hidden theory must be large. Moreover,
as the optimal ``quality" of the BZ tuning becomes better with larger rank,
between theories with different large ranks and the same messenger scale, the smallest mass gap belongs to the theory with the largest rank.

There are however exceptions to the destabilization phenomenon.  SU(N) N=4 sYM is such an exception where  turning on vevs for the Cartan scalars, breaks the gauge group in factors (say $SU(N)\to SU(N-M)\times SU(M)$) and the bi-fundamentals $(N,M-N)$ are massive vector multiplets. At energies much lower than the vevs, the IR theories are N=4 sYM with gauge groups $ SU(M)$ and $SU(N-M)$, namely CFTs.

This exception however relies crucially on the fact that the messengers link two very special CFTs. For a generic non-conformal theory like the SM it seems a miracle that both the messager and SM quantum effects on the running in the hidden CFT cancel out. The only conceivable way out is that the effect of messengers or the SM quantum effect is negligible because the rank of the hidden group is large. The larger the rank the smallest the influence of messengers is on the hidden gauge group.

Consider now the (approximate) CFT with the largest rank (gauge group $\gn$) and large-N matter that we
will not specify at the moment.

The effective degrees of freedom are colorless
glueballs as well as potentially mesons (baryons are very heavy at large N).
Among the effective low-energy degrees of freedom there is always
a spin-two massless state (that is generated from the gauge theory vacuum by the
total stress tensor of theory).
This theory (and N=4 sYM) avoids the Weinberg-Witten non-go theorem because such a state is part of the continuum.

There are other universal composites.
The leading lowest-dimension  operators, that are expected to create glueballs
out of the vacuum are a scalar (the ``dilaton")
$\phi\to Tr[F_{\m\n}F^{\m\n}]$, a spin-2 ``graviton" $g_{\m\n}\to
tr[F_{\m\r}{F^{\r}}_{\n}-{1\over 4}\delta_{\m\n}F_{\r\s}F^{\r\s}]$
and a pseudoscalar ``axion" $a\to
\e^{\m\\r\s}Tr[F_{\m\n}F_{\r\s}]$.
These particles will be massive, and their interactions at low energy are non-perturbative from the
point of view of the gauge theory.

Finally, other adjoint degrees of freedom will provide of a variety of new glueball-like states and give rise to  extra internal dimensions, as standard AdS/CFT indicates.

We summarize this discussion as follows: The dominant hidden theory that is coupled to the SM, should be (approximately) conformal, at strong coupling and of the largest possible rank. We will denote this theory henceforth by $CFT_0$. The messengers coupling it to the SM should be heavy (compared to SM scales).

\section{The IR landscape}

The low energy view of the UV landscape is essentially filtered by the couplings to the SM. We first summarize the three distinct possibilities.
 \begin{enumerate}

\item The SM is completely decoupled from any other theory, but it is completed in the UV to a UV complete theory. This is a non-generic case and although possible it seems highly unlikely.

 \item The SM is coupled to other gauge group factors with small rank via ``messenger" fields of finite mass (necessarily larger than current limits)
This is the conventional definition  case of a hidden sector and such occurrences are plausible.

 \item The SM is coupled to group factors with large rank, via messenger fields that have large masses. This is generically expected. Integrating out the messengers, will induce low energy interactions of the standard model fields to the gauge invariant composites of the large rank theory, namely the lowest spin two-glueball (metric), scalars (dilaton or axion glueballs) and other bound-states.

 \end{enumerate}

Therefore a generic low-energy consequence is that the SM will be coupled to a metric and other fields that comprise a ``gravitational sector".
The couplings of the glueball to leading order are what one expects from diffeomorphism invariance.

The way this can be implemented by renormalizable interactions is  important and needs discussion. The interactions should be renormalizable  if there is only one step and one set of messengers that couples the hidden theory to the SM. There are more complicated setups where there is an intermediate group $G_i$ and two sets of messenger fields. The $(G_N,G_i)$ messengers and the $(G_i,SM)$ messengers. Strong interactions of the $G_i$ group create at energies lower that the characteristic scale $\Lambda_i$, bound states with the quantum numbers of $(G_N,SM)$ messengers. In that case the coupling of the SM fields to such composite messengers does not need to be of the renormalizable type.  We will not entertain further this possibility as it seems more fine tuned without extra gains in the IR. Therefore we assume that the couplings of the messenger field to the SM are renormalizable.

For this to happen, it is important that all SM fields can be written as bi-fundamentals of the SM simple group factors. The possible ways of writing the SM content as bi-fundamentals has been classified in the effort to embed the SM in orientifold vacua of string theory, \cite{class}. A representative simple example, first described in the first reference of \cite{class}, is given in the table below, where the U(1) of the U(3) group (gauged baryon number)  is a very weak gauge interaction and where the SM hypercharge
is identified with $Y={1\over 6}Q_3-{1\over 2}Q_1$.
In the table $V$ stands for fundamental, $A$ for two-index antisymmetric, $S$ for two index symmetric and the gauge bosons have been omitted being adjoints of the respecting groups.

\begin{center}
\begin{tabular}{|c|c|c|c|}
\hline
{ particle}&{ ${U(3)}_c$}&{ ${SU(2)}_w$}&{ ${U(1)}$}\\
\hline
$$&$$&$$&$$\\
\hline
$Q(\b3,\b2,+\frac{1}{6})$&$V$&$ V$&$0$\\
\hline
$U^c(\bb3,\b1,-\frac{2}{3})$&$\bar V$&$0$&$V$\\
\hline
$D^c(\bb3,\b1,+\frac{1}{3})$&$\bar V$&$0$&$\bar V$\\
\hline
$L\;(\b1,\b2,-\frac{1}{2})$&$0$&$\bar V$&$V$\\
\hline
$e^c(\b1,\b1,+1)$&$0$&$0$&$\bar S$\\
\hline
$\nu^R(\b1,\b1,0)$&$0$&$A$&$0$\\
\hline
$H(\b1,\b2,-\frac{1}{2})$&$0$&$\bar V$&$V$\\
\hline
\end{tabular}

\end{center}

Once the SM fields can be written as bifundamentals under the SM group factors then the messenger fields need to be bifundamentals under the large-N group $G_N$ and
one of the factors of the SM groups. Both massive fermions $\chi^a_i$ and vectors $A^{a,i}_{\m}$ are necessary as messagers in order to induce couplings of the metric and other composites to all SM fields\footnote{$a$ is a color index of the large-N group, while $i$ is a fundamental index of one of the SM factor groups.}.
For example a quark can be coupled by a gauge invariant coupling of the form $\bar q^{ij}\gamma^{\mu}\chi^a_{i}A_{\mu}^{a,j}$ to the messagers, while a SM gauge field $B_{\m}^{ij}$ as $\bar\chi^a_{i}\gamma^{\mu}\chi^a_{j}B_{\m}^{ij}$. Finally if the SM Higgs is a fundamental scalar (not composite), then it couples as  $\bar\chi^a_{i}\chi^a_{j}H^{ij}$.  We  may also replace some or all of the massive vectors $A^{a,i}_{\m}$ by scalars $Q^{a,i}$.
Integrating out the messenger fields will generate an effective action for the SM fields coupled to the graviton and other glueballs.

The associated concept of dynamical
space-time is emergent. It  appears after integrating out the messenger fields that average over color degrees of freedom. At large rank this may be thought of as a coarse-graining transformation, responsible for the emergence of thermodynamic properties in the interactions of gravity with matter.

\subsection{Interlude: Anomalies and extra U(1)'s}

It was shown in \cite{class} that all possible ways of writing the SM as a theory with bifundamental fields only involve the presence of at least one, and typically more U(1) gauge factors, beyond what we have in the SM.
It is therefore relevant to expect that there might be new physics associated with such a presence.

 An important issue with the extra U(1)s in the SM sector is that generically the SM spectrum is anomalous with respect to them. The only exception is that the U(1) is proportional to B-L (and there are no extra (massive fields) beyond those of the SM that are charged under this U(1)).

In orientifolds of string theory such ``anomalous U(1)'s are massive, and the gauge boson masses can be lighter than the string scale, \cite{akr}, their masses being suppressed by a one-loop factor compared to string states.
If the U(1) descends from higher dimensions then the mass may depend also on compactification volumes, \cite{akr} and the U(1) may be massive even if it has no 4d anomalies, \cite{anasta}.

In the context of orientifolds, the mass is provided by mixing to string theory axion fields, that are responsible for cancelling their anomalies.
In 4d QFT, the general low-energy structure of the effective action related to anomalies was described in \cite{bianchi} generalizing the work of \cite{df}
and finding a non-trivial structure in the presence of extra U(1)'s\footnote{The low energy effective action and gauge boson mixing have been described in \cite{irges}.}.

In 4d QFT the anomaly can be cancelled only if

\begin{itemize}

\item The anomalous U(1) symmetry is broken by the Higgs mechanism.

\item There are fermions, charged under the anomalous U(1), beyond those of the SM that  cancel  the anomaly and have obtained large masses (much larger than SM scales) from the associated Higgs effect.

\end{itemize}

This is an interesting window into potential new physics, except if the extra U(1) is B-L in which case there is no need for additional fermions. One should however needs to explain in this case why the B-L gauge symmetry is spontaneously broken

\section{The geometric view}

\subsection{4d QFT vs string theory}

The AdS/CFT correspondence has taught us that at Large N new emergent dimensions appear. There is therefore a higher-dimensional picture of the
UV-chaotic QFT landscape we are envisaging resembling a hyper-universe of the type motivated and described in \cite{rev2} filled with branes of various sorts and their associated emergent closed and open strings (of which as we shall see there are many different kinds).

The intuition obtained from matrix models and the AdS/CFt correspondence stipulates that emergent dimensions are associated with continuous eigenvalue distributions of independent (not related by symmetries) gauge variant fields.
Moreover, four-dimensional gauge theories can be used to define (dual) string theories in a similar fashion as two-dimensional $\sigma$-models.

In  2d $\s$-models the basic scalar fields are interpreted as the coordinates of a string, and this provides defacto the dimension of space-time. Additional fermionic fields may provide further Grassmann coordinates.
Continuum strings emerge when the $\s$-model is conformally invariant.
The $\alpha'$-expansion of the  $\s$-model can be perturbative when the string is living in the critical dimension. We know of two critical dimensions for strings moving in Minkowski signature: D=26 for bosonic strings, a theory that is perturbatively unstable, and D=10 for closed fermionic strings of which we know three types: type II, type 0 and heterotic.

Once $D$ is smaller or larger than the critical dimension, the $\alpha'$-expansion generically breaks down as (generalized) curvatures become comparable to the string scale. There is however a qualitative difference
between the supercritical and the subcritical case.
In the subcritical case we have many stable vacua, despite strong curvatures.
They correspond to IR fixed points at ``strong coupling".
On the other hand, in the supercritical case we typically  have runaway behavior (in time).

In 4-dimensions we may construct strings out of a  gauge theory with $N\ggg 1$ so that the associated string theory is weakly coupled, \cite{4dstring}.
The string degrees of freedom and observables are determined by the Wilson loops, and their local field expansion is generated by ``glueballs". We define as glueballs here any single trace gauge-invariant observable made our of adjoint fields.

As argued from different points of view in \cite{4dstring, lorentz} and \cite{lee}, we expect in general that the source functional of QFT, properly defined\footnote{The subtle and non-trivial issue is the non-linear realization of local symmetries that include the diffeomorphism invariance as well as gauge symmetries, \cite{lorentz}. Such symmetries are in one-to-one correspondence with associated global symmetries of the QFT.} , to become the generating functional of a dual string theory, where the sources now are dynamical. In \cite{4dstring,lorentz} it was argued that this will arise due to quantum consistency in analogy with what happens in open string theory. In \cite{lee} it was shown constructively that this can arise by integrating out multitrace operators.

Therefore the source functional of CFT becomes a dual string-theory generating functional with dynamical sources.  The holographic experience suggests that the string theory will be weakly curved and weakly coupled when the CFT is at large $N$ and at strong coupling.

\subsection{The coupling of geometry to the SM}

As argued in the previous sector, a messenger sector connects dynamically the SM to the most dominant, large-N, large coupling CFT$_0$.
If the masses of the messengers sector fields are $M_m$, then at energies $E\ll M_m$ we may integrate out the bifundamental messengers to obtain an effective ``low energy" action that directly couples CFT$_0$ operators to the SM operators. Such couplings are of the multitrace type as each of the two factors (CFT$_0$ and SM) has its own gauge group, and therefore the operators must be singlets under each gauge group.

We will discuss here the coupling of a special operator of CFT$_0$, namely the stress tensor, $T^{CFT}_{\m\n}$. We will ignore for the moment other CFT$_0$ operators like other spin-2 (non-conserved) tensors, vectors and scalars.
These will be discussed later on.

Integrating out the messengers will generate couplings of $T^{CFT}$ to in principle all spin two operators of the SM, $T_i^{\m\n}$ and there are several of them.
The leading form of such coupling is of dimension 8 and is of the form (see appendix \ref{count})
\be
\delta S_T= \sum g_i\int d^4x ~ {T^{CFT}_{\m\n}T_i^{\m\n}\over M_m^4}
\label{1}\ee
However, the fact that the SM, conserves energy and momentum (in the absence of messenger interactions, implies that the coefficients $g_i$ in (\ref{1}) must be such that the coupling is
\be
\delta S_T= \int d^4x ~ {T^{CFT}_{\m\n}T_{SM}^{\m\n}\over M_m^4}
\label{2}\ee
In the case at hand if we introduce a source $h_{\m\n}$ for the stress tensor in CFT$_0$ and integrate out the CFT$_0$ degrees of freedom we will obtain a coupling of the SM to a metric that at the linearized level is obtained by substituting the semiclassical external dimensionless source,
\be
h_{\m\n}\leftrightarrow  {T^{CFT}_{\m\n}\over M_m^4}\;.
\ee
where $T^{CFT}$ is normalized to have a two-point function of order ${\cal O}(N^2)$.

 We therefore obtain the minimal infinitesimal coupling\footnote{Note that as shown in appendix \ref{count}, the source $h_{\m\n}\sim O(N)$.}
\be
\delta S_T= \int d^4x ~ h_{\m\n}T_{SM}^{\m\n}
\label{3}\ee
The non-linear terms conspire to make the transformation properties of $h_{\m\n}$ under a diffeomorphism
\be
x^{\mu}\to x^{\mu}+\epsilon^{\mu}\sp h_{\m\n}\to h_{\m\n}+\nabla_{\m}\epsilon_{\n}+ \nabla_{\m}\epsilon_{\m}
\label{8}\ee
where the covariant derivative is defined with respect to the $h_{\m\n}$ metric.

The conclusion is that this coupling, induced by messengers, is the gravitational minimal coupling of the standard model to the stress tensor of $CFT_0$. The equivalence principle, with respect to this coupling is guaranteed at the non-linear level by overall energy conservation, as argued above.

 \subsection{Holography for CFT$_0$ and the cutoff AdS$_5$ picture}.

 We now turn to the 5-dimensional\footnote{We assume there are no further adjoint symmetries that will give dimensions higher than five in the holographic description. Our generic expectations on the nature of $CFT_0$
warrant this expectation.} holographic language, where the $CFT_0$ operators will be described using gravitational fields in 5 dimensions.

There are several differences from the standard story:
\begin{itemize}
\item The gravitational description of CFT$_0$ is expected to be valid at all scales. This will give rise to an AdS$_5$ bulk spacetime, with metric
    \be
    ds^2={\ell^2\over r^2}(dr^2+dx_{\m}dx^{\mu})\;.
 \label{9}   \ee

\item The coupling of this space-time to the SM, exists only for energies $E\ll M_m$. for higher energies, the messengers must be integrated in, and the physics of the SM coupling to CFT$_0$  can not anymore be described in terms of a standard gravitational coupling.

\item For $E\ll M_m$ we can sketchily describe the gravitational coupling to the SM as
    \be
S=M^{3}\int d^5 x\sqrt{\hat g}\left[R^{(5)}(\hat g)+{12\over \ell^2}\right]+\delta\left(r-{1\over \mu}\right)\int d^4x\sqrt{g}{\cal L}_{SM}(\psi_i,g_{\m\n})
\label{4}\ee
where $M$ is the 5d Planck scale, $\hat g_{\m\n}$ is the 5 dimensional metric, ${\cal L}_{SM}$ is the SM Lagrangian, $\psi_i$ denotes collectively the SM fields and $g_{\m\n}$ is related to the 4d induced metric $g_{\m\n}^{ind}$ on a slice at $r=1/\mu$, and up to factors of O(1), by
 \be
 g_{\m\n}=(M\ell)^{3\over 2}~g^{ind}_{\m\n}
 \ee
 where $r$ is the usual AdS$_5$ coordinate and $\mu$ is the RG energy scale of the SM ($\mu\ll M_m$).
As expected
\be
(M\ell)^3=\kappa ~N^2
\label{10}\ee
with $\kappa$ a number of order one, that depends on the nature of CFT$_0$.

\end{itemize}

To summarize, the setup reminds the picture of cutoff AdS$_5$, appropriate for the Randall-Sundrum context, \cite{apr} with the main difference that AdS$_5$ has no UV cutoff here and the gravitational coupling of the SM model ceases to exist as such near or above $\mu=M_m$.
In that sense the SM can be thought of, in the geometrical picture as a brane located at $r={1\over \mu}$ and coupled to the bulk geometry.

It is important to mention that the setup of the coupling of the bulk fields (sources of the CFT$_0$) to the SM resembles the coupling to a SM brane positioned at the cutoff slice of AdS$_5$. This intuition is very useful as it
helps in estimating several dynamical phenomena by using the dual brane picture.

This is very similar to the picture of the standard  model we have in string theory orientifolds, when the bulk geometry is close to AdS.

\subsection{Non-conserved spin-2 operators}

CFT$_0$ will contain an infinite number of spin-two operators. Only one is expected to be conserved, the stress tensor. This is equivalent to the fact that $CFT_0$ is not a tensor product. Such operators will have anomalous dimensions , so that $\Delta=4+\delta$ with $\delta \sim {\cal O}(1)$, and their couplings to the standard model will be suppressed by an extra power of $M_m^{\delta}$. The vevs of such operators are expected to be negligible as they are irrelevant. The only potentially worrisome case is the case where $\delta $ is infinitesimal. However for this to happen the CFT$_0$ must be a product of two weakly interacting CFTs, \cite{aharony},  that is not expected to be generically the case. This protects the equivalence principle in the SM.

\section{5d vs 4d gravity}

What we have argued so far is that at energies well below the messenger scale the SM interacts with a 5-dimensional metric, the avatar of CFT$_0$.
If this was the whole story, we would have a gross contradiction with experimental data: observable gravity is 4d at the scales where it has been tested.

There is however an effect, first described in geometrical form in \cite{dgp},  namely brane induced gravity, that is relevant here.
Indeed, loop effects of the standard model fields are expected to generate a kinetic term for the fours dimensional metric in (\ref{4}). As from the point of view of the bulk AdS the SM fields are localized in the $r=1/\mu$ slice, this kinetic term with be four dimensional. Moreover, the SM loops, coupled to the metric, have a natural UV cutoff, namely the messenger mass $M_m$.

This does not mean that there are no shorter distance contributions, but that these are not interpretable in terms of a correction to the gravity coupling.
Indeed, the proper way to think of this is to summarize the short distance contributions ($E>M_m$) of SM loops as corrections to the messenger action and its couplings, leading to an effective action at $E=M_{m}$. For much lower energies we use the gravitational picture and continue to compute the lower energy contributions of the SM loops, that now have a natural cutoff $\sim M_m$.

Therefore we can write
  \be
S=M^{3}\int d^5 x\sqrt{\hat g}\left[R^{(5)}(\hat g)+{12\over \ell^2}\right]+\delta\left(r-{1\over \mu}\right)S_4
\label{5}\ee
\be
S_4=\int d^4x\sqrt{g}\left[\Lambda_4+M_4^2 R^{(4)}(g)+\zeta (R^{(4)})^2+\cdots +{\cal L}_{SM}(\psi_i,g_{\m\n})\right]
\label{6}\ee
with
\be
\Lambda_4\sim N^2~M_m^4\sp M_4\sim N~M_m\sp \zeta\sim N^2\log{M_m^2\over \mu^2}\label{666}\ee
 according to the discussion above and the estimates of appendix \ref{count}.

This setup is somewhat different that DGP because of the presence of $AdS_5$, but this is the case studied\footnote{See section 6 of \cite{rev} for a review of brane induced gravity and its interaction with holography.}   in \cite{irs}.

The effective cosmological constant as seen from the gravitational evolution in the SM has been estimated in \cite{kr} to be
  \be
  \Lambda_{eff}=\Lambda_4-24{M^3\over \ell}
  \ee
  To describe the physics of the SM today it is appropriate that a tuning must take place between brane and bulk vacuum energies so that the SM is nearly flat. This should presumably happen dynamically\footnote{Several past efforts have tried to arrange this in RS related geometries but the solutions have been plagued by bad (non-resolvable) singularities. This is expected to change in IR confining geometries, but this problem needs further study.}.
In this case we will have $\Lambda_{eff}\simeq 0$.
However we should also allow the possibility of a large effective cosmological for the purposes of cosmology.
We will introduce the variable $S$ and the 54-4d crossover length scale $r_c$ as follows
  \be
S=\left({\Lambda_4\ell\over24 {M^3}}\right)^{1\over 4}\sp r_c={M_4^2\over M^3}
\label{7}\ee
When $S=1$, then $\Lambda_{eff}=0$.
When $S\gg 1$ then there is a large positive cosmological constant visible in the SM.

We can then estimate as follows
\be
{M_m^4\over M^4}\sim {1\over N^2~M\ell}\sim {S^4\over N^{{8\over 3}}}\sp {M_4\over M}\sim S~N^{1\over 3}
\label{7a} \ee
so that we can relate all relevant scales to the AdS curvature scale
\be
M\sim N^{2\over 3}{1\over \ell}\sp M_m\sim S~ {1\over \ell}\sp M_4\sim N~S~{1\over \ell}\sp r_c\sim S^{2}{\ell}\sp \Lambda_4\sim \left(S~N^{1\over 2}{1\over \ell}\right)^4
\label{7b}\ee
For $S\gg 1$ the effective 4d cosmological constant is $\sim \Lambda_4$
and therefore the effective deSitter curvature length $\ell_4$ is given by
\be
\ell_4=\sqrt{M_4\over \Lambda_4}\sim {\ell\over S}
\label{7c}\ee

To investigate the effective gravitational interaction that emerges for the SM fields we temporarily neglect $\Lambda_{eff}$.
In appendix \ref{b} we analyze a more general case, that of a bulk field, with both a 5d and a 4d kinetic term, and both a 5d and a 4d mass.
We compute the propagator on that field at the position of the SM brane"

The Fourier transform of the graviton propagator is now given by\footnote{Although the bulk fields have transparent boundary conditions at the position of the ``SM brane" and Dirichlet at the AdS boundary, the boundary conditions for the calculation of the propagator of bulk fields on the SM brane should be reflective (as in the RS model) because the brane position is a cutoff for the SM interactions. The general case including bulk and boundary masses is detailed in appendix \ref{b}.} , \cite{irs}
\be
G(\vec p)={1\over p}~{K_2(pr_0)\over M^3 (2K_1(pr_0)+r_cp K_2(pr_0))}
\label{11}\ee
where $p\equiv |\vec p|$, $r_0$ is the radial position of the brane (that is $r_0\sim {1\over M_m}$) and $K_n$ are the standard Bessel functions.

There are two distinct cases in general,
\begin{enumerate}

\item ${r_c\over r_0}\gg 1$. In this case
\be
G(\vec p)\simeq {1\over M_4^2 p^2}
\label{12}\ee
at all $p$, and this is a four-dimensional behavior for the gravitational force of the SM, with a four-dimensional Planck scale $M_4$.

\item ${r_c\over r_0}\ll 1$. In this case
\be
G(\vec p)~~\simeq~~ \left\{ \begin{array}{lll}
\displaystyle {1\over (M_4^2+M^3r_0)p^2},&\phantom{aaa} &pr_0\ll 1,\\ \\
\frac{1}{2M^3p},&\phantom{aaa}&1\ll pr_0\ll {r_0\over r_c},
\\ \\
{1\over M_4^2p^2},&\phantom{aaa}&p\gg {1\over r_c}\;.
\end{array}\right.
\label{13}\ee
and SM Gravity is 4-dimensional at short and long scales ($pr_0\ll 1$ and $p\gg {M^3\over M_4^2}$) with Planck scale $M_4$, and five dimensional at intermediate scales with Planck scale $M$.

\end{enumerate}

Moreover, there is also the UV cutoff in this picture, namely $\Lambda_{UV}\simeq M_m$ that must be considered in the system above, as beyond that scale the gravitational interaction must be replaced by its ancestor, the messenger interactions.
The three length scales that determine the physics of the gravitational interaction according to the above are ${\ell}$, $r_c$, and ${1\over M_m}$ and eventually, if there is a sizable cosmological constant, the scale $\ell_4$.

If we assume that $\Lambda_{eff}\simeq 0$ then $S\simeq 1$ and from (\ref{10}), (\ref{666})  and (\ref{7b}) we obtain
\be
 {1\over M_m}\sim r_c \sim {\ell}\sp M_P\equiv \sqrt{M_4^2+M^3r_0}\sim {N\over \ell}\gg {1\over \ell}
\label{14}\ee
so that we are in the first case above and indeed $M_P\simeq NM_m\simeq M_4$. Therefore the gravitational interaction is four-dimensional at all scales and the messenger mass determines directly the four-dimensional Planck scale.

Interestingly, all scales are of the same order suggesting that the 5d origin of the gravitational interaction, starts becoming apparent only at the cutoff scale where the interaction is not any more a good description of the underlying messager interactions.
On the other hand, the four-dimensional Planck  scale, is hierarchically larger (at large N) from the cutoff of the gravitational picture.

We should note that this state of affairs (that the cutoff is a factor N lower than the Planck scale) was argued from entanglement entropy considerations in \cite{dvali}.

In the un-tuned (deSitter) case, $S\gg 1$ we obtain instead
\be
r_c\gg {\ell}\ll {1\over M_m}\sim {\ell_4}
\ee
Therefore all transition phenomena are masked in this case by the presence of the effective SM cosmological constant, while at these scales the gravitational interaction is 4d.

\subsection{Non-minimal Higgs-gravity couplings}

There is a single relevant (as opposed to marginally relevant) coupling in the SM, namely the Higgs mass term. SM loops with two external Higgs lines and a graviton will generate to leading order two more terms,
\be
\delta S_{one-loop}=\int d^4x\sqrt{g}\left[M_H^2+\zeta'R^{(4)}\right]|H|^2
\label{15}\ee
where $H$ is the Higgs doublet of the SM and
\be
M_H\sim N M_m\sim M_P\sp \zeta'\sim N^2\log{M_m^2\over \mu^2}\;.
\ee
The first term is the standard term associated with the hierarchy problem.
It should either be fine tuned with the tree level term or it will not appear if the theory can handle the hierarchy problem. We will not worry about this further here, but this is something that requires further investigation.

The second has been advocated to drive inflation at early times provided $\zeta'$ is large enough, \cite{shapo}.
 At large N this is indeed the case, and what was advocated in \cite{shapo} becomes natural.

\section{The SM equivalence principle revisited}

So far we have considered the coupling of spin-2 operators of CFT$_0$ with those of SM. We have argued that they lead to standard gravity (by combining holography and DGP ideas), and the the SM equivalence principle remains intact.

There are of course other operators from CFT$_0$ that can couple to the SM, and such operators are a serious threat to the equivalence principle.

\subsection{Scalar operators}

The most generic kind that we will address in this section are scalar operators that are parity-even. The most generic such operator in a gauge theory in four dimensions is $tr[F^2]$, dual to the string theory dilaton. There can be many others. Their dimensions can span the whole range of relevant values $1\leq \Delta \leq 4$ as well as be irrelevant with $\Delta\geq 4$.

We have already argued that RG stability of $CFT_0$ suggests that there are no relevant scalar operators and we will therefore assume so.
Irrelevant operators (being massive in the bulk holographic picture) are not expected to influence as backgrounds the SM even if they are coupled to it.

The potential danger to the equivalence principle arises from nearly marginal or marginal CP-even scalar operators are they can have couplings to the SM operators (after integrating out the messengers) of the  form
\be
\delta S_{\rm scalar}\sim \int d^4x\sqrt{g}\left[\sum_i c_i {O(x)O_i(x)\over M_m^4}+ c_H {O(x)|H(x)^2|\over M_m^2}+\cdots\right]
\label{16}\ee
Here $c_i,c_H$ are dimensionless coefficients, $O$ is the scalar marginal operator from CFT$_0$ (normalized as in appendix \ref{count}), and $O_i$ are (nearly) marginal CP-even scalar operators in the SM.
Replacing   the dimensionless field
\be
\Phi\leftrightarrow N{O\over M_m^4}
\label{17}\ee with a source for the CFT$_0$ operator  the action becomes
\be
\delta S_{\rm scalar}\sim N\int d^4x\sqrt{g}\left[\sum_i c_i \Phi ~O_i+ c_H {M_m^2 \Phi |H^2|}+\cdots\right]
\label{18}\ee
It is clear that such couplings destroy the equivalence principle at ${\cal O}(1)$ level, as well as the provide new worrisome couplings to the Higgs.

However, as in the gravity case, the DGP phenomena come to the rescue.
SM Loops will generate both a 4d kinetic term and a 4d mass for the scalar $\Phi$, that must be added to its 5d bulk kinetic terms.
From holography we expect to have
\be\delta S_{\Phi}=M^3\int d^5 x\sqrt{\hat g}\left[(\partial \Phi)^2+V(\Phi)\right]+N^2~\delta\left(r-{1\over \mu}\right)\int d^4x\sqrt{g}\left[(\partial\Phi)^2+m^2\Phi^2+\cdots\right]
\label{19}\ee
with $m\sim M_m$. The 5d potential depends on the choice of the operator and the CFT. Its minimum is at $\Phi=0$.  If the scalar $\Phi$ is a marginal operator, then it  is massless around the minimum. Its further structure was studied in \cite{ihqcd} and further refined in \cite{bourdier}.   Our conclusions on the nature of the linearized interactions mediated by $\Phi$ are insensitive to the rest of the details of $V(\Phi)$.  However, such a potential may be important for the cosmology of the SM.

Unlike the gravitational case, the 4d mass of the scalar $\Phi$ generated is of the order of the cutoff scale $M_m$ (which is hierarchically lower than the Planck scale).
Therefore, such a field is not visible in the low energy theory, neither it mediates any important long range interaction. When it becomes important, and therefore violates the equivalence principle,  the gravity description is no longer valid.

In the gravity case, 4-dimensional diffeomorphism invariance has prohibited the generation of a  mass for the four dimensional graviton. Only the fifth component has effectively obtained a mass from the SM loops, \cite{dgp}.

We conclude that such operators are barred from low energy SM physics and are in particular not a direct danger for the equivalence principle.
However , it is fair to say that since the equivalence principle is valid a a very high accuracy, the arguments above need refinement in order to understand in more detail both the accuracy required, and where new physics may be visible.

  \section{Pseudoscalar operators and axions}

There is one special pseudoscalar operator in CFT$_0$ that requires a separate treatment. The reason is that, like the metric and unlike the rest of the scalar operators, it is protected by symmetries.

The operator in question is the (pseudoscalar) instanton density\footnote{If CFT$_0$ has a product gauge group, there will be several such axion operators, one for each simple factor of the gauge group. We will not examine this possibility here.}
\be
A(x)\equiv Tr[F\wedge F]\;.
\label{ax} \ee
 In a gauge theory, its source is known as the $\theta$ angle and it is renormalized only non-perturbatively.
In the classical theory we have the Peccei-Quinn symmetry $\theta\to \theta+constant$. Semiclassically, this continuous symmetry is broken to a discrete $Z$ symmetry $\theta\to \theta+2\pi$ due to instantons.

The holographic dual field of $A$ is a string theory axion, $a$. In the canonical example of $N=4$ sYM, it is the RR axion of IIB string theory.
There are several peculiarities in the bulk holographic action for the axion.

\begin{itemize}

\item At large $N$, the axion kinetic term is suppressed by an additional power of $1/N^2$ compared with the rest of the tree level terms. This can be understood from the fact that the axion is a RR field, and this fact matches properly, \cite{diss} the Witten-Veneziano solution to the $\eta$'-mass problem,  \cite{witten}.

\item The classical Peccei-Quinn symmetry translates into a translation symmetry for the bulk axion to all orders in $1/N$ expansion, \cite{ihqcd}. Therefore there is no axion potential, but the glueballs and $\eta'$ have a mass gap.

\item D-instantons (which are the holographic dual of the CFT$_0$ instantons) generate an axion potential that breaks the Peccei-Quin symmetry to a discrete one.
In the sequel  we will neglect this potential as it suppressed non-perturbatively at large N as ${\cal O}(e^{-N})$.

\end{itemize}

In the bulk effective action,  the axion mixes with other scalar fields. Since we argued above that such scalar fields acquire 4-d masses of order $M_p$, we will consider them as non-fluctuating (and with zero vevs) and we will neglect such couplings.
We may therefore write for the 5d axion action respecting the Peccei-Quinn symmetry
\be
S_{axion,5d}={M^3\over N^2}\int d^5x \sqrt{\hat g}\left[~{Z\over 2}(\partial a)^2+{\cal O}((\partial a)^4)\right]
\label{20}\ee
where we explicitly indicated the large-N suppression factor.   The constant normalization factor $Z\sim {\cal O}(1)$ is important,  if we normalize $a$ so that its exact symmetry after instantons is $a\to a+1$. In large-N YM, it determines the topological susceptibility, \cite{ihqcd,ax}.

To what extend this axion can mediate a new long-range force for the SM fields depends on its couplings to them. They can be determined in the QFT, by explicitly integrating out the messengers.

The PQ symmetry implies that all couplings to SM fields are derivative couplings with one exception: the messenger sector has an analogue of the axial $U(1)_A$ symmetry for its fermions. This symmetry will be generically  anomalous in terms of the gauge group of the CFT$_0$, and therefore the ``messenger $\eta'$" will mix with the CFT$_0$ instanton density $A$.
 A (generic) cubic anomaly between the SM gauge symmetries and this messenger axial symmetry implies that the axion will couple to the appropriate SM instanton densities.
Define
\be
Tr[T^{a}_iT^b_i T_{U(1)}]_{\rm messenger}=N~I_i
\label{23} \ee
 where $i=1,2,3$ denotes the three gauge factor of the SM, $T^a_i$ are the generators of the SM gauge group in the messenger representation, $T_{U(1)}$ is the generator of the axial U(1) symmetry of the messengers. We have pulled out a factor of $N$ as the messengers are fundamentals of the CFT$_0$ gauge group, and this color factor acts as multiplicity in the trace in (\ref{23}). Therefore $I_i$ if not zero, is an ${\cal O}(1)$  number.

In view of this,  the 4d action for the axion operator $A(x)$ in (\ref{ax}) (after integrating out the messengers) contains
\be
S_{axion,4d}=\sum_{i=1}^{3}\int d^4x~ {A(x)\over M_m^4} ~I_i~Tr[F_i\wedge F_i]+{1\over N}\int d^4x\sqrt{g}\left[ {|H|^2(\pa A)^2\over M_m^6}+{(\pa A)^2 Tr[F_i^2]\over M_m^8}+\cdots\right]
\label{24} \ee
The terms included in the dots above  are of the type $(\pa A)^2 \bar \psi i\slashed{D} \psi$, $(\pa A)^2 \bar \psi \psi H$ and higher order and they are suppressed by the messenger mass squared at least.
In the ${\cal O}(N^{-1})$ part of (\ref{24}) all coefficients of ${\cal O}(1)$ were set to 1 for simplicity.

In the holographic picture,
we must replace the axion operator $A(x)$ with its source $a(x)$ which should be roughly
\be
a\sim {A\over M_m^4}\sim {Tr[F\wedge F]\over M_m^4}
\ee
and  is dimensionless.

the axion action contains both the  5d part in (\ref{20})  as well as the 4d part in (\ref{24})
\be
S_{axion}=S_{axion,5d}+\delta\left(r-{1\over \mu}\right)\tilde S_{axion,4d}
\label{21}\ee
with
\be
\tilde S_{axion,4d}=\sum_{i=1}^{3}\int d^4x~ a ~I_i~Tr[F_i\wedge F_i]+{1\over N}\int d^4x\sqrt{g}\left[ M_m^2{|H|^2(\pa a)^2}+{(\pa a)^2 Tr[F_i^2]}+\cdots\right]
\ee

The DGP effects will also generate a 4d kinetic term  for the axion localized at the SM slice. Unlike the generic scalar case, no mass or potential term can be generated by SM loops as this is prohibited by the PQ symmetry.
This can be checked explicitly using diagrammatics.
Therefore, after including such loop effects the final axion action is
 \be
S_{axion}={M^3\over N^2}\int d^5x \sqrt{\hat g}~{Z\over 2}(\partial a)^2+
\delta\left(r-{1\over \mu}\right)\int d^4x\sqrt{g}\left({f^2\over 2}+\xi |H|^2+\cdots\right)(\pa a)^2+
\label{25}\ee
$$+\sum_{i=1}^{3}\int d^4x~ a ~I_i~Tr[F_i\wedge F_i]
$$
Again the size of the kinetic term  is controlled by the cutoff, namely
\be
f\sim M_m\sim {M_P\over N}
\label{22}\ee
Finally, to this action, the contributions of the SM Instantons should be included. The most important ones are those of QCD (if $I_{QCD}\not =0$) and they will add to the effective axion action a potential that is\footnote{If $I_{QCD}=0$ then the axion potential will be generated by electro-weak instantons and its scale will be many decades of orders of magnitude lower.}
\be
\delta S_{axion}=\Lambda_{QCD}^4\int d^4\sqrt{g}\cos a
\label{26}\ee
This gives a 4d mass to the axion $\sim \Lambda_{QCD}$ that is hierarchically smaller than the 4d masses of all other marginal scalars that are $\sim \Lambda_{UV}\sim M_m\gg \Lambda_{QCD}$.

As for the interactions mediated by the axion field, the calculation is identical to the one for gravity\footnote{The difference is the presence of the (small) mass of the axion due to instanton effects.}.
In particular, the axion interaction is turned off below the 4d axion mass, and it is four-dimensional at all other scales up to the UV cutoff $M_m$.

\section{Vector operators}

CFT$_0$ may have vector operators. As already argued earlier and shown in appendix B, operators that correspond to  massive fields in  the bulk decouple from the SM Physics. For a vector, to be massless in the bulk, it must be exactly conserved. Therefore the case that remains to discuss is gauge fields  dual to possible exact global symmetries of CFT$_0$.

Consider such a U(1) 5d gauge field $A_{\m}$ that we will call the ``graviphoton" with a bulk action
 \be
 S_{5d}\sim M^3\int \sqrt{g}F^2\sp F_{\m\n}=\pa_{\m}A_{\nu}-\pa_{\n}A_{\m}
\label{27}\ee
A crucial property, that is decisive for the effective couplings of $A_{\m}$ to the SM is whether the messengers are minimally charged  under such a symmetry.
Experience from string theory indicates that that cannot happen\footnote{In string theory the analogue of a U(1) symmetry coming from the hidden CFT is a graviphoton from the closed string sector. The messenger sector is realized by fields on a D-brane. D-brane fields can never be minimally charged under bulk gauge fields in string theory. }
The only exception are flavor-type symmetries of the hidden CFT but these do
not a gravitational profile.

An important next question is whether there are anomalies that couple this symmetry current to SM gauge fields.
We first consider U(1)s where  such anomalies are absent, ie $Tr[Q Q_iQ_i]=0$
with $Q$ the U(1) generator and $Q_i$ SM gauge generators.
Then all couplings between $A_{m}$ and the SM must be strictly gauge invariant. This implies that there are no minimal couplings.

At the linearized level in $A_{\m}$ all such couplings are classified by gauge-invariant antisymmetric tensor operators of the SM.
Such operators have been classified in \cite{lorentz} as they are the most relevant operators breaking Lorentz invariance, and they give rise to topological  currents.

In the standard model the lowest dimension such coupling is between $A_{\m}$ and the hypercharge gauge boson $B_{\m}$ with field strength $G_{\m\n}$.
Other relevant/marginal  gauge invariant antisymmetric operators are
$\Lambda_{\m\n}=\bar\psi\gamma_{\m\n}\psi$, $\hat \Lambda_{\m\n}=\pa_{\m}H\partial_{\n}H^{\dagger}-(\mu\leftrightarrow \n)$
\be
S_{mix}=\int d^4x~F_{\m\n}\left[s_1~G^{\m\n}+s_2 \Lambda^{\m\n}+s_3\hat\Lambda^{\m\n}+\cdots\right]
\label{28}\ee
where as usual $s_i$ are coefficients of ${\cal O}(N)$.
Note that $\pa^{\m}G_{\m\n}$,  $\pa^{\m}\Lambda_{\m\n}$, $\pa^{\m}\hat \Lambda_{\m\n}$ are topological conserved currents, and therefore the graviphoton couplings preserve the graviphoton gauge invariance in the quantum theory.

SM Quantum effects are subtle in this case and we will leave this issue for further work.

We comment below on the two remaining cases: Anomalous U(1)'s and non-abelian
global symmetries.
The characteristic example of a anomalous global symmetry involves a non-trivial  trace $Tr[Q Q_iQ_i]$.
In this case beyond the couplings in (\ref{28}) there is an extra contribution from a triangle graph with one insertion of the global current and two SM gauge fields will generate a generalized Chern-Simons term\footnote{The SM group factor can also be abelian.}, \cite{bianchi}
\be
S'_{mix}=Tr[Q Q_iQ_i]\int d^4x A\wedge CS_3 \sp CS_3=Tr[W\wedge F_W+{2\over 3}W^3]
\label{29}\ee
Note that the closure of the Chern-Simons term is responsible for this action is gauge invariant under the U(1) Gauge transformations.

Finally, in the cases of a a non-abelian gauge symmetry, none of the terms in (\ref{28}) is gauge invariant. One should have at least a quadratic term in $F$, either the scalar  $Tr[F^2]$ or the symmetric tensor $Tr[F_{\m\n}^2]$, which will couple to scalar or symmetric tensor operators of the standard model. In all cases the coupling with be quadratic in the source, and therefore the interactions that will be induced suppressed by an extra power of the Planck scale.

\section{Discussion and open questions}

We have analyzed basic consequences of our three assumptions in section 2, for gravity and other universal interactions and their coupling to the SM. A main assumption was a notion of genericity, that we did not make very precise (and is probably hard to do so without a concrete overview of 4d CFTs).

What  we have concluded is as follows
\begin{enumerate}

\item From all sectors of the UV CFT that the SM is coupled to, the one that will dominate the IR is the theory that is
    \begin{itemize}

    \item (a) a CFT.

    \item (b) Has the maximal number of colors N.

    \item (c) Is at strong coupling.

    \end{itemize}
Therefore this theory, named CFT$_0$, has a weakly curved holographic dual, that we have assumed that lives in 5 dimensions.

\item The presence of a messenger sector of masses $M_m$ implies that CFT$_0$ and the SM will be coupled gravitationally at energy scales below $M_m$ (after integrating out the messengers).
    In particular the gravitational couplings will be universal, but other operators of the CFT$_0$ will also be coupled spoiling a priori the equivalence principle.
Moreover the induced gravitational interactions are five dimensional.

\item Quantum effects of the SM fields alter (locally in energy) the effective action for the gravitational sector. In a holographically dual language, they induce interaction terms that are 4d, and are localized in the position of the ``SM brane" inside the 5d bulk.

\item The result of such 4d terms above is that gravity now becomes 4d, and generic scalar and tensor operators obtain masses that render them (almost) harmless for the equivalence principle.

\item There is a special pseudo-scalar operator, the instanton density of CFT$_0$ that is protected by the topological symmetry. Its dual field is an axion, that has both 5d and 4d interactions turned-on. The second
include standard axion-type interactions with SM instanton densities that are determined by anomalies of the overall theory. Moreover, unlike generic scalar operators, the axion's 4d mass term  is not determined by the messenger mass $M_m$ but by SM instanton effects.

\end{enumerate}

This framework generates a set of scales that have some natural hierarchies.
\begin{itemize}

\item  First the holographic dual  of CFT$_0$ has a 5d Planck scale $M$ and an AdS$_5$ curvature scale $\ell$. They are unobservable in CFT$_0$ and only $(M\ell)^2\sim N^2$ is an observable.

\item The messenger mass $M_m$ is an important input in this picture and determines the 4d Planck scale as $M_P\sim NM_m$, and the induced
    4d cosmological constant, $\Lambda_4\sim N^2M_m^4$.

\item The effective 4d cosmological constant $\Lambda_{eff}=\Lambda_4-24{K^3\over \ell}$ maybe anywhere from zero in the tuned case, to a large and positive is t he 4d contribution dominates the 5d one.

\item Scalars and other tensors that violate the equivalence principle have 4d masses of order $M_m$ and couplings that are of gravitational strength. The masses therefore control the degree of the braking of the equivalence principle.

\item The axion has an axion scale $f\sim M_m\sim {M_P\over N}$.

\item Finally the dimensionless coupling of the curvature to the Higgs is controlled by $N^2$.

\end{itemize}
At the end, two parameters, $N$ and $M_m$ control the physics, and one combination is fixed by the 4d Planck scale, $M_P\sim NM_m$.

There are several interesting and open questions:

\begin{enumerate}

\item  Can $N$ be fixed or measured ?

\item How cosmology,  as observed by the SM,  fits in this picture. One possible answer is given by the brane picture of this setup. If the SM is view as a brane embedded in the 5d bulk, the natural end-point is that it will ``fall" in the bulk, \cite{mirage}.
    The equation of motion that determines this motion is determined both by the bulk background (AdS) as well as the couplings to the SM.
In particular this setup allows for issues associated with the cosmological constant and early-time inflation to be addressed by using the gravitational scalars, or maybe the axion as inflatons.
In particular, there should be a universal CP-even scalar associated with the string theory dilaton that can considered as a natural candidate for the inflaton.

It is interesting that in this context, the masses of such gravitational scalars are always hierarchically smaller than the 4d Planck scale.

The above,  on the other hand,  does not exclude the possibility that the SM Higgs plays the role of the inflaton.

\item More detailed calculations are necessary in order to assert with more precision the couplings of CFT$_0$ operators, especially non-universal ones to the SM, making in such a case more detailed estimates of the violations of the equivalence principle.

\item The case of graviphotons, their couplings to the SM as well as their quantum generated 4d effective action due to SM quantum effects, requires further analysis. This  is the most dangerous case of equivalence principle violation, but it seems that in this case the graviphoton interaction will remain 5-dimensional.

\item The slightly extended form of the SM, required in order to be of the bifundamental type must be explored further. In particular the role of additional U(1)'s must be analyzed and their phenomenological consequences/constraints studied.

 \item We have assumed a hidden CFT at all scales. This assumption may be relaxed in two directions. The first is to assume that the hidden theory is a CFT at and below the messenger scale. All the arguments made in this paper hold unchanged. However, the leading irrelevant operator in that hidden theory may generate non-trivial and interesting corrections.

     The other option to assume that the theory is a CFT down to incredible small scales. Such theories can be certainly engineered by generalizing the Banks-Zaks setup, \cite{4dstring}.
     One could therefore naturally reproduce a theory that is a CFT for many decades of e-folding in energy, and become a gapped theory in the far IR with a characteristic scale $\Lambda=m~e^{-N}$ where $m$ is the mass scale the breaks conformal invariance, (quark mass in BZ), and the number $N$ a very large number. This situation will generate a graviton with a tiny mass, as advocated earlier in \cite{rev}.

 \item The presence of other large N, large couplings CFTs that may be coupled to standard model introduces other metrics coupling to the standard model. There is always an overall metric that is diff-invariant, \cite{lorentz} but in the rest, gravitons become massive \cite{aharony}.
     It is interesting to study such masses and how they may affect the gravity observable in the standard model.

\end{enumerate}

We plan to address these questions in the near future.

\vskip .2cm
\section{\bf Acknowledgments}

\noindent

I am indebted to many colleagues overs the last ten years (too many to remember) for discussions, ideas and criticism on topics related to the issues discussed here. Their enlightening discussions helped and the potential errors here are mine. I would like to thank all of them.

This work was supported in part by European Union's Seventh Framework Programme under grant agreements (FP7-REGPOT-2012-2013-1) no 316165,
PIF-GA-2011-300984, the EU program ``Thales'' MIS 375734, by the European Commission under the ERC Advanced Grant BSMOXFORD 228169 and was also co-financed by the European Union (European Social Fund, ESF) and Greek national funds through the Operational Program ``Education and Lifelong Learning'' of the National Strategic Reference Framework (NSRF) under ``Funding of proposals that have received a positive evaluation in the 3rd and 4th Call of ERC Grant Schemes''. I also thank the ESF network Holograv for support.

 \newpage
\appendix

 \renewcommand{\theequation}{\thesection.\arabic{equation}}
\addcontentsline{toc}{section}{Appendices}
\section*{APPENDIX}

\section{Large-N counting\label{count}}

In this appendix we will derive the large-N estimates of various calculations and effective couplings that are used in the main text.

For simplicity we will consider an adjoint SU(N) action of the matrix model form to do the large-N counting,
\be
S\sim {1\over g^2}\tr\left[MM^{\dagger}+M M^{\dagger}MM^{\dagger}+\cdots\right]
\label{b31}\ee
with propagator
\be
\langle M_{ij}M^{*}_{kl}\rangle\sim g^2\delta_{ik}\delta_{jl}
\label{b32}\ee
Define the single trace operator $O\equiv Tr[MM^{\dagger}]$ and $\lambda\equiv g^2N$ to obtain
\be
\langle O\rangle\sim \la N\sp \langle OO\rangle_{c}\sim \lambda^2\sp  \langle OOO\rangle_{c}\sim {\lambda^3\over N}
\label{b33}\ee
while with the normalization $\tilde O\equiv {N\over \lambda}O$ we obtain
\be
\langle \tilde O\rangle\sim N^2\sp \langle \tilde O\tilde O\rangle_{c}\sim N^2\sp  \langle \tilde O\tilde O\tilde O\rangle_{c}\sim {N^2}
\label{b34}\ee
The operator $O$ has the group structure of all the interesting operators we consider in this paper. Similar results hold for higher single trace operators like $Tr[MM^{\dagger}MM^{\dagger}]/g^4$, where only the powers of $\lambda$ change in the formulae above.

The standard holographic action for $\phi$ dual to $O$ is written as
\be
S\sim M^3\int d^5x\sqrt{g}\left[ (\pa \phi)^2+\tilde g\phi^3 +\cdots\right]
\label{b35}\ee
where we write $M^3=\bar M^3 ~N^2$ to make the $N$ dependence of the 5d Planck scale explicit.
A holographic calculation of the correlation functions, using (\ref{b35})  gives
\be
\langle \phi\phi\rangle\sim \langle \phi\phi\phi\rangle\sim \cdots\sim N^2
\label{b36}\ee
because the correlators are proportional to derivatives of the action in (\ref{b35}) (which is of order ${\cal O}(N^2)$) with respect to the sources.

Therefore,  we have the correspondence
\be
\phi \leftrightarrow \tilde O\sim NTr[MM^{\dagger}]
\label{b37}\ee
However, for the axion that is dual to the $\theta$ angle, with bulk action of O(1)
\be
S_a=\bar M^3\left[\int (\pa a)^2 +{1\over N^2}(\pa a)^4+\cdots \right]
\label{b38}\ee
we have
\be
a\leftrightarrow  Tr[F\wedge F]
\label{b39}\ee

The messenger action can be written schematically as
\be
S_m=\sum_{i,j=1}^N \sum_{\a=1}^{\hat N}(M_{ij}q^{\a}_i\tilde q^{\a}_j+M_{ij}^*\tilde q_i^{\a}q_j^{\a})+\sum_{\a,\b=1}^{\hat N}\sum_{i=1}^N (A_{\a\b}q^{\a}_i\tilde q^{\b}_i+A_{\a\b}^*\tilde q^{\a}_iq^{\b}_i)\sp \langle q_i^{\a}\tilde q^{\n}_{j}\rangle=\delta _{ij}\delta^{\a\b}
\label{b40}\ee
with $q^{\a}_i,\tilde q^{\a}_i$ being the messenger fields, and where we have indicated  their color index with respect to the hidden CFT group ($i,j$) and under the SM group ($\a,\b$).

Consider now the calculation of terms in the action mixing the CFT operator $Tr[MM^{\dagger}]$ with SM fields (say field strengths $F_{\a\b}$).
To do this we first calculate the expectation value of 
\be
V_{1,1}= Tr[MM^{\dagger}]Tr[F_{\a\b}F^*_{\a\b}]
\ee
 in the messenger theory.

To leading order, it is given in the large-N perturbation theory by
\be
\langle V_{1,1}\rangle \sim \langle Tr[MM^{\dagger}](qM\tilde q)(\tilde q M^{\dagger}q)(q A\tilde q)(\tilde q A^{\dagger} q)Tr[F_{\a\b}F^*_{\a\b}]\rangle
\label{b46}\ee
 The relevant diagram is a  box with external insertions of two $M$ operators and two $F_i$ operators.
After contracting the $M,A$ fields it boils down to
\be
\langle V_{1,1}\rangle\sim g^4 \langle (q^{\a}_k\tilde q^{\b}_k)(q^{\b}_l\tilde q^{\a}_l )(q^{\g}_i  \tilde q^{\g}_j)(  q_j^{\d}\tilde q^{\delta}_i)\rangle
\label{b47}\ee
The contractions must now be done so that we obtain a connected diagram, and therefore
\be
\langle V_{1,1}\rangle\sim g^4N^2\sim {\cal O}(1)
\label{b48}\ee
The result is of the same order as the two point function of the $Tr[MM^{\dagger}]$
operator as integrating out the messenger fields generates effectively an extra $Tr[MM^{\dagger}]$ in the expectation value.

To calculate now the coupling $\xi_{1,1} Tr[MM^{\dagger}]Tr[F_{\a\b}F^*_{\a\b}]$ in the effective theory, we must use it to calculate $V_{1,1}$ and match with th messenger result
\be
\langle V_{1,1}\rangle\simeq \xi_{1,1}\langle Tr[MM^{\dagger}]Tr[MM^{\dagger}]Tr[F_{\a\b}F^*_{\a\b}]Tr[F_{\a\b}F^*_{\a\b}]\rangle\sim {\cal O}(1)
\label{b49}\ee
(\ref{b49}) and (\ref{b48}) imply that 
\be
\xi_{1,1}\sim {\cal O}(1).
\ee

We can also calculate
\be
\langle V_{2,1}\rangle= \langle Tr[MM^{\dagger}]Tr[MM^{\dagger}]Tr[F_{\a\b}F^*_{\a\b}]\rangle
\ee
We obtain
\be
\langle V_{2,1}\rangle\sim g^8 \langle (q^{\a}_k\tilde q^{\b}_k)(q^{\b}_l\tilde q^{\a}_l )(q^{\g}_i  \tilde q^{\g}_j)(  q_j^{\d}\tilde q^{\delta}_i) (q^{\a'}_m  \tilde q^{\a'}_n)(  q_n^{\b'}\tilde q^{\b'}_m)\rangle\sim g^8N^3\sim {\cal O}\left({1\over N}\right)
\label{b47}\ee
Again this implies a coupling of the form $\xi_{2,1}Tr[MM^{\dagger}]^2Tr[F_{\a\b}F^*_{\a\b}]$ in the effective theory. Similarly, $\xi_{2,1}$ can be estimated from the effective theory correlator
\be
\langle V_{2,1}\rangle\simeq \xi_{2,1}\langle Tr[MM^{\dagger}]^4~Tr[F_{\a\b}F^*_{\a\b}]^2\rangle\sim {\cal O}(1)
\ee
from where we deduce that 
\be
\xi_{2,1}\sim {\cal O}\left({1\over N}\right)
\ee

We may therefore write:
\be
S_{eff}=S_{CFT_0}+S_{SM}+\int d^4x ~{O(x)P(x)\over M_{m}^{\Delta_O+\Delta_P-4}}+{\cal O}\left({1\over N}\int d^4x ~O(x)^2P(x)\right)
\label{b41}\ee
where $O(x)$ is a single trace operator of CFT$_0$ with dimension $\Delta_O$, with the normalization as in (\ref{b33}) and $P(x)$ is a gauge-invariant operator of the SM of dimension $\Delta_{P}$ . We have also suppressed constants of order one, at large $N$.
The powers of the messenger mass $M_m$ are dictated by scaling dimensions and the fact that this mass is the only scale that enters the calculation\footnote{The full quantum field theory is assumed to be UV complete (renormalizable) and therefore the full result can be written in terms of only the {\em renormalized} $M_m$ defined precisely at the decoupling scale.}.

We will now convert this to a coupling between the holographic version of CFT$_0$ and the SM.
This can be obtained by adding the source dual to $O$ and using the AdS/CFT rule:  the 5d on-shell action $S_5(\phi_0)$ for a 5d scalar $\phi$ with a source $\phi_0(x)$ should be equal to $W(\phi_0)$,  the generating functional defined as
\be
e^{-W(\phi_0(x))}\equiv \langle e^{N\int d^4x ~\phi_0(x)~O(x)}\rangle_{CFT_0}\;.
\label{b43}\ee
For small $\phi_0$ we have
\be
W(\phi_0)=N^2\int d^4x d^4y\left[\langle O(x)O(y)\rangle \phi_0(x) \phi_0(y)+{\cal O}(\phi_0^3)\right]
\label{b44}\ee
If we now add to (\ref{b41}) a source term as in (\ref{b43}) the path integral over CFT$_0$ will give $W$ evaluated at
 $$\phi_0\to \phi_0+{P(x)\over NM_m^{\Delta_P+\Delta_O-4}}\;.$$
The only difference here is that this coupling exists below a given radial scale $r\sim {1\over M_m}$ as above it one has to integrate-in the messengers.
Therefore the term in the 5d effective action that mixes $\phi_0$ to $P(x)$ is obtained from (\ref{b43}) as
\be
\delta W={N\over M_m^{\Delta_P+\Delta_O-4}}\int_{|x-y|>{1\over M_m}} d^4x d^4y\left[ {\phi_0(x) P(y)\over |x-y|^{2\Delta_O}}+\cdots\right]
\label{b45}\ee
$$
\simeq{N\over M_m^{\Delta_P+\Delta_O-4}} \int d^4x ~\phi_0(x) \int_{|x-y|>{1\over M_m}} d^4y \left[ { P(x)\over |x-y|^{2\Delta_O}}+{\square P(x)\over |x-y|^{2\Delta_O-2}}+\cdots\right]
$$
$$
\simeq{N\over M_m^{\Delta_P+\Delta_O-4}} \int d^4x \left[\phi_0(x)P(x)\int_{|y|>{1\over M_m}}{d^4y\over |y|^{2\Delta_O}}+ \phi_0(x)\square P(x)\int_{|y|>{1\over M_m}}{d^4y\over |y|^{2\Delta_O-2}}+ \cdots\right]
$$
$$
\simeq{N\over M_m^{\Delta_P+\Delta_O-4}} \int d^4x \left[c_0{\phi_0(x)P(x)\over M_m^{4-2\Delta_O}}+c_1{\phi_0(x)\square P(x) \over M_m^{6-2\Delta_O}}+ \cdots\right]
$$
$$
\simeq {c_0~N\over M_m^{\Delta_P-\Delta_O}}\int d^4x ~\phi_0(x)P(x)+{c_2~N\over M_m^{\Delta_P-\Delta_O+2}}\int d^4x ~\phi_0(x)\square P(x)+\cdots
$$
where in the second line we expanded $P(y)$ around $x$ in the third line we shifted $y\to y+x$ and in the fourth line the coefficients $c_n$ are ${\cal O}(1)$ and stem from the calculation of the integrals.

For the YM axion $a$ that instead is dual to $Tr[F\wedge F]$, the mixing  to the SM Instanton densities $Tr[F_i\wedge F_i]$ along the same lines is
\be
S\sim {Tr[F\wedge F]Tr[F_i\wedge F_i]\over M_{m}^4}\sim ~a ~Tr[F_i\wedge F_i]
\label{b42}\ee

\vskip 1cm

\section{Interactions of bulk semilocalized fields in AdS\label{b}}

We generalize in this appendix the calculation of \cite{irs} by including both bulk and brane masses  for bulk fields. We will consider a scalar, but the results hold more generally. The effects of higher curvature corrections were analyzed in \cite{ktt}.

Consider the action for a bulk scalar $\Phi$
\be
S={1\over 2}M^3\int d^5 x \sqrt{g_5}((\pa\Phi)^2+M_5^2\Phi^2)+M^3r_c{\ell\over 2r_0}\delta(r-r_0)\int d^4x\sqrt{g_4}((\pa\Phi)^2+m^2\Phi^2)
\label{a1}\ee
Note that the large-N estimates we have given in the body of the paper imply that
\be
M^3\sim N^2\sp r_c\sim {\cal O}(1)\sp m\sim M_m\sim {1\over r_0}
\label{a10}\ee

Going to  Fourier space for the brane coordinates, we obtain for the propagator $G(p,r)$
\be
\left[ \pa_{r}^2-{3\over r}{\partial\over \pa r}-\left(p^2+{M_5^2\ell^2\over r^2}\right)\right]G(p,r)-r_c\left(p^2+{m^2\ell^2\over r_0^2}\right)\delta(r-r_0)G(p,r)=-{1\over M^3}\delta(r-r_0)
\label{a2}\ee
The propagator must have reflective (RS) boundary conditions at $r=r_0$ that implies that $\lim_{ r\to r_0^+} {\pa_r G}=-\lim_{ r\to r_0^-}{\pa_r G}$

We may solve (\ref{a2}) away from $r=r_0$ to obtain
\be
G(p,r)=B(p) r^2 K_{\nu}(pr)\sp r>r_0\sp \nu =\sqrt{4+M_5^2\ell^2}
\label{a3}\ee

Integrating (\ref{a2}) around $r=r_0$ we obtain the discontinuity equation for the derivative
\be
\pa_r G\Big|_--\pa_r G\Big|_+=r_c\left(p^2+{m^2\ell^2\over r_0^2}\right)G(p,r_0)-{1\over M^3}
\label{a4}\ee
from which we can obtain the normalization of the propagator
\be
B={1\over M^3}{1\over \left[-4r_0+r_c(p^2r_0^2+m^2\ell^2)\right]K_{\nu}(pr_0)-2r_0^2p K'_{\nu}(pr_0)}
\label{a5}\ee

From (\ref{a3}) we can evaluate the propagator on the brane
\be
G(p,r_0)={1\over M^3}{r_0^2 K_{\nu}(pr_0)\over \left[-4r_0+r_c(p^2r_0^2+m^2\ell^2)\right]K_{\nu}(pr_0)-2r_0^2p K'_{\nu}(pr_0)}
\label{a6}\ee
$$
={1\over M^3}{r_0 \over 2\nu-4+{r_c\over r_0}(p^2r_0^2+m^2\ell^2)+2pr_0 {K_{\nu-1}(pr_0)\over K_{\nu}(pr_0)}}
$$
where we used
\be
K'_{\nu}(x)=-{\nu\over x}K_{\n}(x)-K_{\nu-1}(x)
\label{a7}\ee
The following dimensionless scales affect the behavior of the propagator: $pr_0$, $m\ell$, ${r_c\over r_0}$, $\nu$.

To proceed further we need the asymptotics of the Bessel functions. For $x\to 0$ we obtain
\be
{xK_{\nu-1}(x)\over K_{\nu}(x)}={x^2\over 2(\nu-1)}+{\cal O}(x^4)
\label{a8}\ee
while for $x\to \infty$ we obtain
\be
{xK_{\nu-1}(x)\over K_{\nu}(x)}=x-{2\nu-1\over 2}+{\cal O}\left({1\over x}\right)
\label{a9}\ee

Note that constant part of the propagator in (\ref{a6}) $2\nu-4+{r_c\over r_0}m^2\ell^2$ obtains contributions from the 5d and 4d masses.
The 5d contribution $2\nu-4$ is positive for irrelevant operators of the CFT$_0$, zero for marginal operators, and negative for relevant operators.
Using the estimates of (\ref{a10}) we obtain for the 4d (dimensionless)  mass term
\be
{r_c\over r_0}m^2 \ell^2\sim (M_m\ell)^5
\label{a11}\ee
and in particular to be ${\cal O}(1)$ in the large N expansion.

Note that in our case $r_0$ is associated with the messenger mass and therefore $pr_0 \ll 1$ as $r_0$ is essentially the UV cutoff for the gravitational description. It is the  (\ref{a8}) asymptotics that is relevant and the propagator becomes
\be
 G(p,r_0)={1\over M^3}{r_0 \over 2(\nu-2)+{r_c\over r_0}(p^2r_0^2+m^2\ell^2)+{(pr_0)^2\over \nu-1}}
\label{a12}\ee
and can be written as
\be
 G(p,r_0)^{-1}=M^3\left({r_0\over \nu-1}+r_c\right)\left[p^2+M_{eff}^2\right]\sp M_{eff}^2=(\nu-1){2(\nu-2)+{r_c\over r_0}m^2\ell^2\over r_0((\nu-1)r_c+r_0)}
\label{a13}\ee
From (\ref{a13}) we can observe that the interaction of $\Phi$ on the SM  brane is 4d, with mass $M_{eff}$ and with a Planck scale
\be
M_P^2=M^3\left({r_0\over \nu-1}+r_c\right)
\label{a14}\ee
The ratio
\be
{r_c\over r_0}\sim M_m {N^2 M^2_m\over M^3}\sim N^2 {M_m^3\over M^3}\sim S^3
\label{a15}\ee
controls which of the two summands dominates the Planck scale.
In the last step in (\ref{a15}) we have used (\ref{7b}).
It therefore clear that $r_c$ dominates $r_0$, and at best it can become of the same order. Therefore for order of magnitude estimates we can set $r_c\simeq r_0$.

For the graviton, $\nu=2$, $m=0$ and
\be
M_P^2=N^2M_m^2\left(1+{1\over S^3}\right)
\label{a16}\ee
In the fine tuned case $S=1$ but in the deSitter case $S$ can be large and its contribution to the Planck scale becomes negligible.

Irrelevant operators, $\nu>2$, even if we neglect the 4d mass , have an effective mass
\be
M_{eff}^2\sim {1\over r_0r_c}\simeq {M_m^2\over S^3}
\label{a17}\ee
In the nearly flat case, $S\simeq 1$ their effective mass is the cutoff
but in the deSitter region, $S\gg 1$ it may become much smaller.

For generic scalars with $m\sim M_m$ the contribution to the effective mass is of the order
\be
{m^2\ell^2\over r_0^2}\sim M_m^4\ell^2\sim M_m^2 S^2
\label{a18}\ee
It is clear from (\ref{a17}) and (\ref{a18}) that for the flat case, $S\simeq 1$, the two contributions to the mass are of the same order.
However in the deSitter regime $S\gg 1$, the 4d mass dominates and becomes much larger than the cutoff.

Finally for the axion, the 4d instanton-induced mass $m\simeq \lambda_{QCD}\ll M_m$. Since the axion is nearly marginal the only contribution to its mass comes from 4d. The effective mas for its interaction becomes
\be
M_{a,eff}\sim S~\Lambda_{QCD}
\ee
and can be enhanced in a deSitter setup.

\newpage


\end{document}